\documentclass[aps,nofootinbib,amsfonts,superscriptaddress]{revtex4}
\usepackage{amsfonts,amssymb,amscd,amsmath}
\usepackage{graphicx}
\usepackage{subfigure}
\usepackage{bm}
\graphicspath{{figures/}}

\begin{document}

\begin{flushright}
LU TP 16-45\\
August 2016
\vskip1cm
\end{flushright}

\title{Drell-Yan process in $pA$ collisions: the exact treatment of coherence effects}

\author{Victor P.  Goncalves}
\email{victor.goncalves@thep.lu.se}
\affiliation{Department of Astronomy and Theoretical Physics, Lund University,
SE-223 62 Lund, Sweden}
\affiliation{High and Medium Energy Group, Instituto de F\'{\i}sica e
Matem\'atica, Universidade Federal de Pelotas, Pelotas, RS, 96010-900, Brazil}

\author{Michal Krelina}
\email{michal.krelina@fjfi.cvut.cz}
\affiliation{Czech Technical University in Prague, FNSPE, B\v rehov\'a 7, 11519
Prague, Czech Republic}

\author{Jan Nemchik}
\email{nemcik@saske.sk}
\affiliation{Czech Technical University in Prague, FNSPE, B\v rehov\'a 7, 11519
Prague, Czech Republic}
\affiliation{Institute of Experimental Physics SAS, Watsonova 47, 04001 Ko\v
sice, Slovakia}

\author{Roman Pasechnik}
\email{roman.pasechnik@thep.lu.se}
\affiliation{Department of Astronomy and Theoretical Physics, Lund University,
SE-223 62 Lund, Sweden}

\begin{abstract}
In this work, we investigate production of Drell-Yan (DY) pairs in proton-nucleus collisions 
in kinematic regions where the corresponding coherence length does not exceed the nuclear radius, $R_A$, 
and the quantum coherence effects should be treated with a special care. The results for the nucleus-to-nucleon 
production ratio available in the literature so far are usually based on the assumption of a very long coherence 
length (LCL) $l_c\gg R_A$. Since the onset of coherence effects is controlled by the coherence length $l_c$, 
we estimated its magnitude in various kinematic regions of the DY process and found that the LCL approximation 
should not be used at small and medium c.m. collision energies ($\sqrt{s} \lesssim 200$ GeV) as well as at large dilepton 
invariant masses. In order to obtain realistic predictions, we computed for the first time the DY cross section using the generalised 
color dipole approach based on the rigorous Green function formalism, which naturally incorporates the color transparency and 
quantum coherence effects and hence allows to estimate the nuclear shadowing with no restrictions on the CL.
In addition to the shadowing effect, we studied a complementary effect of initial state interactions (ISI) that causes 
an additional suppression at large values of the Feynman variable. Numerical results for the nuclear modification factor 
accounting for the ISI effect and the finite $l_c$ are compared to the data available from the fixed-target FNAL measurements 
and a good agreement has been found. Besides, we present new predictions for the nuclear suppression as a function of dilepton 
rapidity and invariant mass in the kinematic regions that can be probed by the RHIC collider as well as by the planned AFTER@LHC and 
LHCb fixed-target experiments. 
\end{abstract}
\maketitle

\section{Introduction}

Hadronic production of massive lepton pairs, known as the Drell-Yan (DY) process, is a clean, precise and controllable probe for short-distance dynamics 
and partonic structure of hadrons (for a recent review see, e.g. Ref.~\cite{peng}). In particular, the DY process on nuclear targets is an ideal tool to probe 
and to quantify the initial state interaction (ISI), saturation, gluon shadowing and coherence effects in a clean environment due to the absence of final-state 
interactions and fragmentation processes typically associated with energy loss or absorption phenomena \cite{BGKNP}. The corresponding predictions for 
the nucleus-to-nucleon ratio known as the nuclear modification factor $R_{pA}$ have been obtained using the color dipole approach \cite{zkl,k95,bhq97,kst99,krt01,
dynuc,rauf,gay,pkp,Basso,Basso_pp} which is known to provide as precise predictions for the DY cross section in $pp$ collisions as the Next-to-Leading-Order (NLO) 
collinear factorisation framework. In the color dipole picture, the DY process off nuclei looks as $\gamma^*$ Bremsstrahlung off a projectile quark propagating 
through the nuclear medium \cite{kst99}. Remarkably, the dipole framework enables us to include the coherence effects in nuclear collisions naturally from 
the first principles by means of the generalised path-integral (or Green function) formulation (see e.g. Ref.~\cite{kst99}), and thorough studies of its implications for 
the DY reaction on nuclear targets is the main purpose of this work.

The onset of nuclear coherence effects (nuclear shadowing) is controlled by the coherence length (CL), $l_c$, which can be interpreted as a lifetime of $\gamma^*$-quark
fluctuations in a nuclear environment. At high energies, the CL exceeds the nuclear radius $R_A$, $l_c \gg R_A$, and the corresponding formalism 
treating the coherence effects is greatly simplified. This regime referred to as the large coherence length (LCL) limit was discussed in details in Ref.~\cite{BGKNP}, 
where predictions for the nuclear suppression of the DY pair production at RHIC and LHC energies were presented.

On the other hand, at small and medium c.m. collision energies ($\sqrt{s} \lesssim 200$ GeV) when the CL is much shorter than the mean nucleon spacing in a nucleus,
i.e. $l_c \lesssim  1\div 2$ fm, the corresponding regime is known as the short coherence length (SCL) limit where no effects of the quantum coherence are expected.
Consequently, the formalism used in Ref.~\cite{BGKNP} and valid in the LCL limit cannot be extended directly to such kinematic region.

An alternative way is to study the DY pair production process using the Green function formalism developed in Ref.~\cite{kst99} which represents a universal method 
to describe dipole interactions with a nuclear target including effects responsible for the nuclear shadowing. This formalism naturally incorporates the color transparency and 
quantum coherence effects and can consequently be used in analysis of the nuclear shadowing without particular restrictions on magnitude of the CL.

So far, the existing calculations of the nucleus-to-nucleon ratio of the DY pair production cross sections are performed in either LCL or SCL limits and do not account 
for a realistic value of the CL. The LCL results systematically overestimate the nuclear shadowing effect suggesting the need for a more sophisticated analysis which 
should incorporate finite CLs. As the main objective of the present paper, we investigate for the first time the DY process in the framework of generalised color dipole 
approach based on the Green function formalism. The latter enables us to obtain realistic predictions for the nuclear shadowing in those kinematical regions where 
the SCL and LCL regimes are not realised. We also verify that this formalism correctly reproduces both the SCL and LCL approximations in the kinematical regions 
corresponding to very short and very long CLs, respectively. 

The paper is organised as follows. In the next Section, we introduce the CL and present its dependences on energy, Feynman variable $x_F$ and invariant dilepton 
mass $M_{l\bar l}$. We have specified the kinematic regions where the DY cross section can be reliably predicted using the SCL and LCL approximations. 
We have also determined the transition domains between the SCL and LCL regimes where the use of the rigorous Green function formalism is unavoidable 
for realistic predictions for the DY nuclear cross section. In Section \ref{formalism}, we present the color dipole description of the DY process and briefly 
discuss the key elements of the Green function formulation. In particular, we describe the numerical method used in our analysis in order to obtain a numerical 
solution of the Schr\"odinger equation for the Green function determining the dipole propagation through a color medium. In Section \ref{res}, we present the results 
for the nuclear modification factor $R_{pA}$ and compare them with the data available from the FNAL fixed-target experiment. The predictions for $R_{pA}$ 
as a function of DY pair rapidity and invariant mass are given in kinematical regions that can be probed by future measurements at RHIC collider as well as 
by the AFTER@LHC and LHCb fixed-target experiments. These calculations are performed for several models of the dipole cross section. At large Feynman 
variable $x_F$ and/or at forward rapidities, besides the quark shadowing inherited from the Green function formalism, we take into account the gluon shadowing (GS) 
contribution as well leading to an additional nuclear suppression. Besides the shadowing corrections, we incorporate yet another source of nuclear effects 
dominated at large dilepton invariant masses and $x_F$ values and caused by the initial-state interactions which are relevant in kinematic regions where no shadowing 
is expected. In order to test the considered Green function formalism, we verify that it reproduces the results obtained in the standard SCL and LCL approximations
in the corresponding kinematic regimes. Finally, in Section \ref{conc} we summarize our main conclusions.

%
\section{Coherence length in the DY process}
%

The color dipole approach is formulated in the target rest frame where the DY pair production process is viewed as $\gamma^*$ Bremsstrahlung off 
a projectile quark (see e.g. Refs.~\cite{Basso_pp,pkp}). In the case of nuclear targets, the DY process is controlled by the CL which regulates the 
interference of scattering amplitudes on different nucleons in the target nucleus. The CL for the $|q\gamma^* \rangle$ fluctuation is given by 
the uncertainty relation
\begin{equation}
\label{eq:dipole:lc-1}
    l_c = \frac{2E_q}{M^2_{q\gamma}}\,,
\end{equation}
where $E_q$ refers to the energy of the projectile quark, and $M_{q\gamma}$ is the effective mass of the considered $|q\gamma^* \rangle$ fluctuation
defined as
\begin{equation}
\label{eq:dipole:qy-mass}
    M^2_{q\gamma} = \frac{M_{l\bar{l}}^2}{1-\alpha} + \frac{m_q^2}{\alpha} + \frac{p_T^2}{\alpha(1-\alpha)}\,.
\end{equation}
Here, $M_{l\bar{l}}$ is the dilepton invariant mass, $m_q$ is the projectile quark mass, $p_T$ is the transverse momentum of the virtual photon, and 
$\alpha$ is the light-cone momentum fraction taken by the photon from the projectile quark. Using the kinematic relations
\begin{eqnarray}
E_q = \frac{x_1}{\alpha}E_p \,, \qquad x_1 x_2 = \frac{M_{l\bar{l}}^2+p_T^2}{s} \,, \qquad s = 2 m_p^2 + 2 E_p m_p \sim 2 E_p m_p \,,
\end{eqnarray} 
where $x_1$ is the light-cone momentum fraction taken by the photon from the incoming proton, $m_p$ is the proton mass, and $E_p$ is the projectile 
proton energy, one obtains the following expression for the CL
\begin{equation}
\label{eq-cl}
  l_c = \frac{1}{x_2 m_p} \frac{(M_{l\bar{l}}^2 + p_T^2)(1-\alpha)}
  { (1-\alpha) M_{l\bar{l}}^2 + \alpha^2 m_q^2 + p_T^2}\equiv \frac{1}{x_2 m_p} K(\alpha,p_T)\,.
\end{equation}
\begin{figure}[t]
\large
\begin{center}
\scalebox{0.5}{\includegraphics{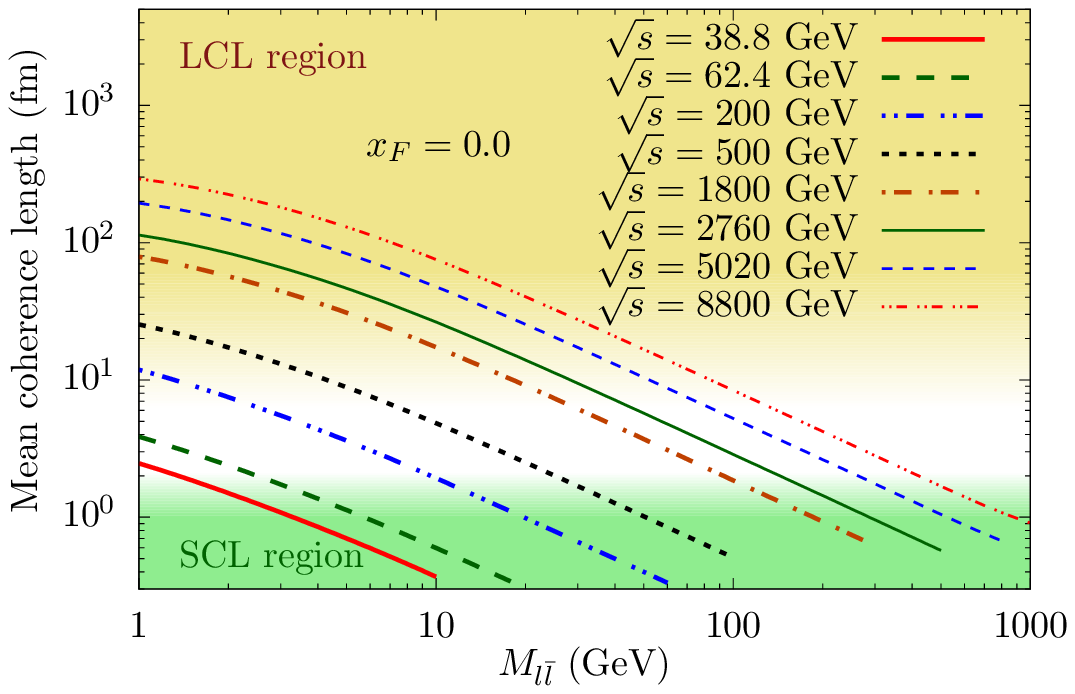}}
\scalebox{0.5}{\includegraphics{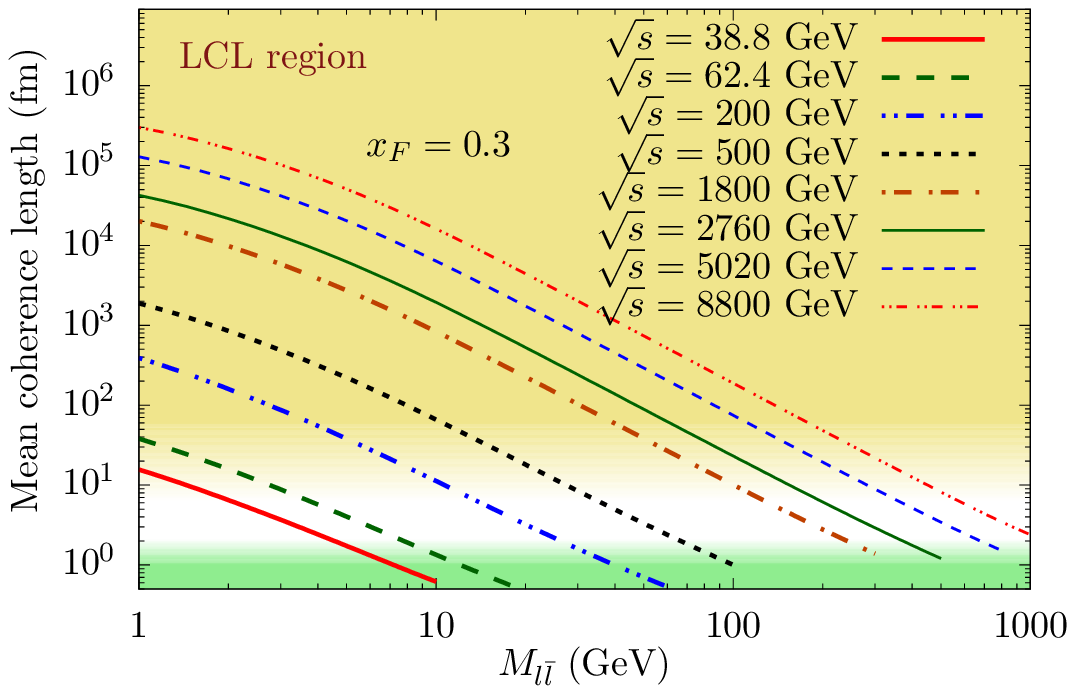}}
\scalebox{0.5}{\includegraphics{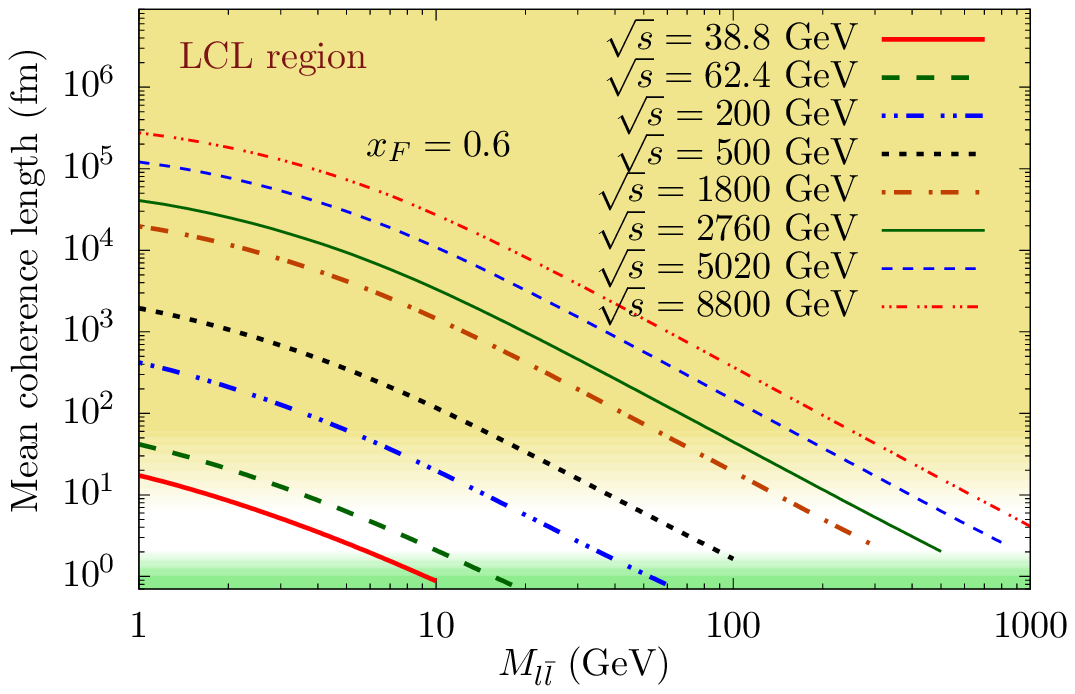}}
\caption{
The mean CL as a function of the invariant dilepton mass for several values of the c.m. collision
energy $\sqrt{s}$ and Feynman $x_F$ variable. The LCL and SCL regions are highlighted by the yellow 
and green bands, respectively.}
\label{fig:lc_M}
\end{center}
\end{figure}
\normalsize
\begin{figure}[t]
\large
\begin{center}
\scalebox{0.5}{\includegraphics{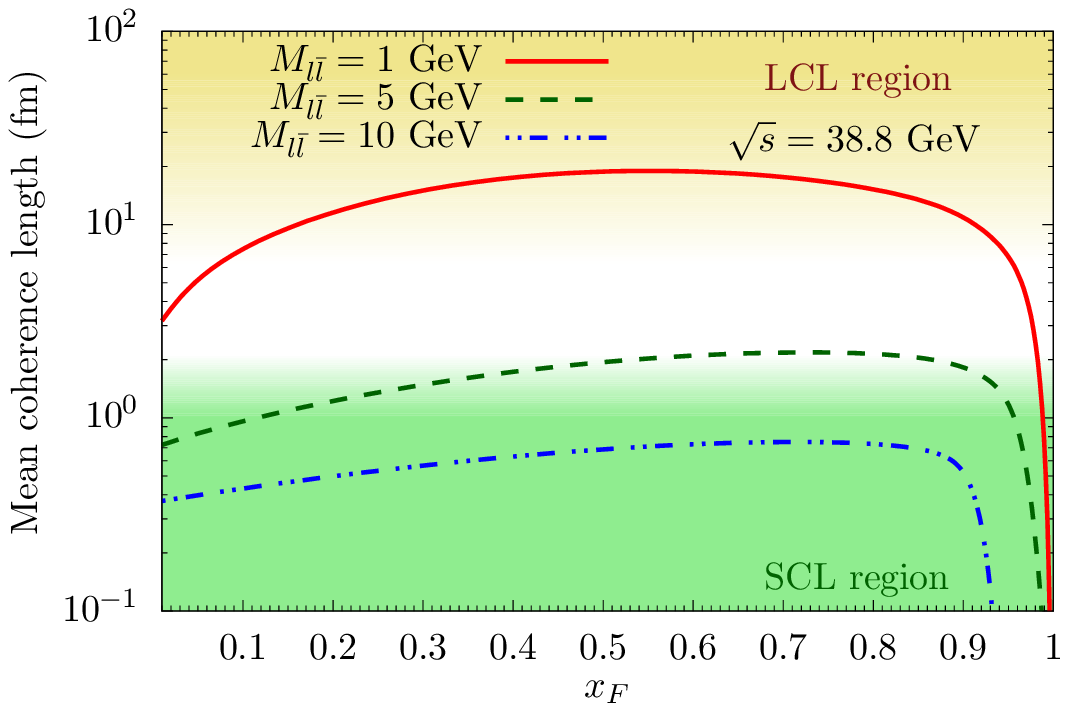}}
\scalebox{0.5}{\includegraphics{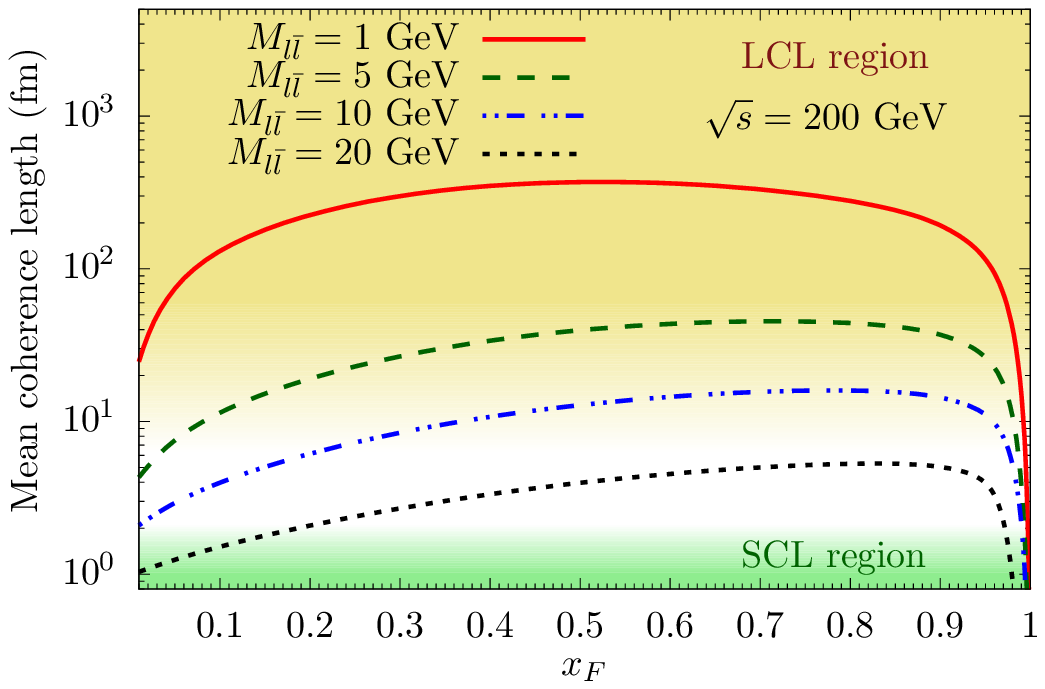}} 
\scalebox{0.5}{\includegraphics{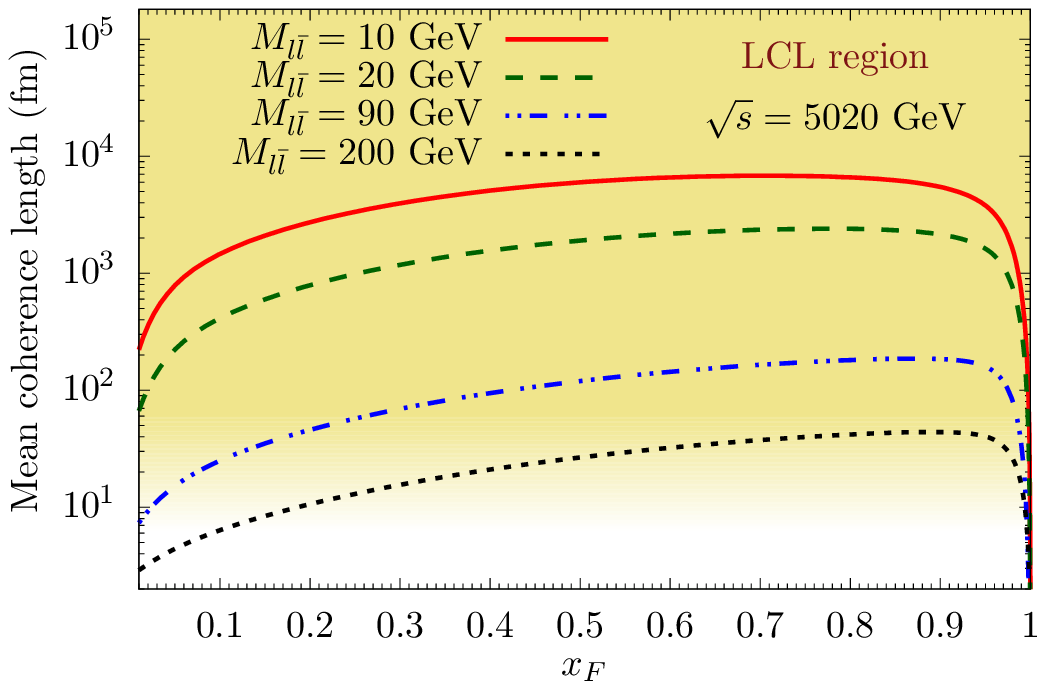}}
\caption{
The mean CL as a function of the Feynman $x_F$ variable for several values of the dilepton invariant mass 
$M_{l\bar{l}}$ and c.m. collision energy $\sqrt{s}$. The LCL and SCL regions are highlighted by the yellow 
and green bands, respectively.}
\label{fig:lc_xF}
\end{center}
\end{figure}
\normalsize

In order to analyse the c.m. energy, Feynman $x_F$ and dilepton invariant mass dependences of the CL, we follow 
the procedure developed in Ref.~\cite{Johnson:2001xfa} such that
\begin{equation}
    \langle l_c \rangle_{\alpha} = \frac{1}{x_2 m_p}\frac
    {\int_{x_1}^1 \frac{d\alpha}{\alpha^2}  
     \int d^2p_T  \sum_q \left( f_q\left(\frac{x_1}{\alpha}\right) 
  + f_{\bar q}\left(\frac{x_1}{\alpha}\right) \right) 
\frac{d^3\sigma^{(qN \to \gamma^* X)}}{d (\ln \alpha) d^2 p_T}K(\alpha,p_T)}
    {\int_{x_1}^1 \frac{d\alpha}{\alpha^2}
     \int d^2p_T  \sum_q \left( f_q\left(\frac{x_1}{\alpha}\right) + f_{\bar q}
    \left(\frac{x_1}{\alpha}\right) \right) 
    \frac{d^3\sigma^{(qN \to \gamma^* X)}}{d (\ln \alpha) d^2 p_T}} \,,
\end{equation}
where $ K(\alpha,p_T)$ is defined in Eq.~(\ref{eq-cl}), $f_q$ ($f_{\bar q}$) represents the starting quark (anti-quark) distribution function 
for a flavour $q$, and the differential cross section corresponds to the quark-nucleon interaction (see below).  

Multiple scatterings off nucleons in the target nucleus give rise to various nuclear effects. One of such effects is the shadowing which represents one of 
the sources for the nuclear attenuation in the DY process. The CL controls the number of scatterings of a projectile in the course of its propagation through 
a nuclear target and, therefore, determines the magnitude of nuclear effects \cite{Johnson:2001xfa,Johnson:2007kt}. One should distinguish between 
the two limiting cases where the physical picture and the corresponding description of nuclear effects in DY pair production is significantly simplified:
\begin{itemize}
\item \textbf{The short CL (SCL) regime}. 
In this case, the mean CL $\langle l_c \rangle_{\alpha}$ becomes much smaller than the mean nucleon (interparticle) spacing in the nucleus, i.e.
$\langle l_c \rangle_{\alpha} < d$ = $1\div2\,$fm. Such a short CL excludes any shadowing since the time scale of the fluctuation is so short that 
the constituents of the fluctuation have no time to multiply interact with the medium. Thus, all nucleons of the nuclear target contribute 
equally to the cross-section. This is the so-called Bethe-Heitler regime \cite{Bethe:1934za}.
\item \textbf{The long CL (LCL) regime}. 
This regime represents the situation when the mean CL considerably exceeds the nuclear radius, i.e. $\langle l_c \rangle_{\alpha} \gg R_A$, such that 
the projectile interacts with the whole nucleus at the surface. This is the so-called Landau-Pomeranchuk-Migdal effect 
\cite{Landau:1953um,Migdal:1956tc} corresponding to the maximal shadowing.
\end{itemize}

The analysis of the transition region between the SCL and LCL regimes is generally much more difficult and less straightforward. The most rigorous 
approach is based on the Green function formalism \cite{Kopeliovich:1998gv,Raufeisen:1998rg} that will be described in the next Section. 
Another and more simple alternative, that sometimes is used in the literature, is based on a simple interpolation between the SCL and LCL limits 
using the longitudinal nuclear form factor $F_A(q_c,b)$, where the variable $q_c = 1 / l_c$ corresponds to the longitudinal momentum 
transferred in the scattering \cite{Johnson:2001xfa,Kopeliovich:2002yh}.

In Fig.~\ref{fig:lc_M} we present our predictions for the mean CL as a function of the dilepton invariant mass $M_{l\bar{l}}$ for several values of 
the c.m. collision energy $\sqrt{s}$ and Feynman $x_F$ variable. Here, the LCL and SCL regimes are highlighted as bands of different colours. 
Note, for central rapidities (i.e. $x_F = 0$) and c.m. collision energies smaller than 200 GeV, the DY process probes essentially the SCL and transition 
(represented by a white band) regions. In contrast, at LHC energies the DY process is controlled only by the LCL regime over a large range of dilepton 
invariant masses. Although at forward rapidities (corresponding to $x_F = 0.6$ in the plot) the LCL domain gets significantly expanded, the SCL and transition 
regions can still be probed at smaller c.m. energies.

In Fig.~\ref{fig:lc_xF} we analyse the $x_F$ dependence of the mean CL $\langle l_c \rangle_{\alpha}$ for different energies and dilepton invariant masses. 
One can see that for $\sqrt{s} = 5020$ GeV, the LCL regime is realised over a large range of invariant masses. On the other hand, for much smaller c.m. energies, 
e.g. $\sqrt{s} = 38.8$ GeV corresponding to the fixed-target FNAL experiment, the DY process is controlled essentially by the SCL and transition regimes. 

At RHIC energy $\sqrt{s} = 200$ GeV, the LCL limit represents a good approximation in description of the nuclear shadowing only for small dilepton invariant 
masses and at forward rapidities as is demonstrated in Fig.~\ref{fig:lc_M}. Note, the $x_F$ (or $x_1$) behaviour of the mean CL differs significantly from 
a simple scaling  $l_c\approx 1/(2m_p x_2)$ at large $x_1 \rightarrow 1$ and $x_F \rightarrow 1$. In this limit, the mean CL vanishes $\langle l_c \rangle_{\alpha} 
\rightarrow 0$ as one notices from Eq.~(\ref{eq-cl}) such that the nuclear effects (shadowing) should vanish as well. Figs.~\ref{fig:lc_M} 
and \ref{fig:lc_xF} suggest that in order to obtain realistic predictions for the DY production cross sections in $pA$ collisions one should employ a generic
framework applicable for any values of the CL.

%
\section{Drell-Yan process in $pA$ collisions: the Green function formalism}
\label{formalism}
%

Assuming not very large values of the dilepton invariant mass $M_{l\bar l}\ll M_Z$, where $M_Z$ is the $Z^0$ boson mass,
it is sufficient to account for Bremsstrahlung of a heavy photon $\gamma^* \to l\bar l$ only. In the color dipole approach
the transverse momentum $p_T$ distribution of the photon Bremsstrahlung in a quark-nucleon interaction reads \cite{kst99}
\begin{equation}
\label{eq-qN-pT}
  \frac{d^3\sigma^{(qN \rightarrow \gamma^* X)}}{ d \ln \alpha d^2 p_T} =
  \frac{1}{(2\pi)^2}
  \int d^2\rho_1 d^2\rho_2
  e^{i \vec p_T \cdot (\rho_1 - \rho_2)}
  \Psi_{\gamma^*q}^\dagger(\alpha,\vec\rho_2) \Psi_{\gamma^*q}(\alpha,\vec\rho_1)
  \Sigma(\alpha, \vec \rho_1, \vec \rho_2)\,,
\end{equation}
where $\Psi_{\gamma^*q}(\alpha,\vec\rho)$ are the light-cone wave functions of the projectile $q\to q\gamma^*$ fluctuation 
(see e.g. Ref.~\cite{Basso_pp}), respectively, and
\begin{equation}
\label{eq-sigma}
  \Sigma(\alpha, \vec \rho_1, \vec \rho_2) = \frac{1}{2} \left( \sigma_{q \bar q}(\alpha\vec \rho_1) + 
  \sigma_{q \bar q}(\alpha\vec \rho_2) - \sigma_{q \bar q}(\alpha(\vec \rho_1 - \vec \rho_2)) \right) \,.
\end{equation}
Here, $\sigma_{q \bar q}(\vec \rho\,)$ denotes the universal dipole cross section which determines interaction of a $q\bar q$ dipole
of transverse separation $\vec \rho$ which a nucleon at high energies (small $x$), first introduced in Ref.~\cite{zkl}. It is a flavor-independent
universal function of $\vec{\rho}$ and energy allowing to describe various high-energy processes in a uniform way.

The production cross section of DY pairs in $pp$ collisions is given by a convolution of the quark-nucleon, $qN\to \gamma^*X$, DY cross section 
with the corresponding parton distribution functions (PDFs) $f_q$ and $\bar f_q$ of the incident hadron, i.e.
\begin{equation}
\label{eq-pp-pT}
   \frac{d^4\sigma^{(pp \rightarrow l^+l^-X)}}{d^2p_T dx_F dM_{l\bar{l}}^2} = 
   \sigma^{(\gamma^* \rightarrow l^+ l^-)} \frac{x_1}{x_1+x_2} \int_{x_1}^1
  \frac{d\alpha}{\alpha^2} \sum_q Z_q \left( f_q(x_1/\alpha,Q^2) + 
  \bar f_q(x_1/\alpha, Q^2) \right) \frac{d^3\sigma^{(qN \rightarrow \gamma^* X)}}
  { d \ln \alpha d^2 p_T} \,,
\end{equation}
where $Z_q$ is the fractional quark charge, $Q^2 = p_T^2+(1-x_1)M_{l\bar{l}}^2$ is the hard scale of the process, and the factor 
$\sigma^{(\gamma^* \rightarrow l^+ l^-)}=\alpha_{em}/3\pi^2$ accounts for the $\gamma^* \to l\bar l$ transition. Integrating Eq.~(\ref{eq-pp-pT}) 
over the dilepton transverse momentum $p_T$, one obtains
\begin{equation}
\label{eq-pp}
  \frac{d^2\sigma^{(pp \rightarrow l^+l^-X)}}{ dx_F dM_{l\bar{l}}^2} = 
  \sigma^{(\gamma^* \rightarrow l^+ l^-)} \frac{x_1}{x_1+x_2} \int_{x_1}^1
  \frac{d\alpha}{\alpha^2} \sum_q Z_q \left( f_q(x_1/\alpha,Q^2) + 
  \bar f_q(x_1/\alpha, Q^2) \right) \frac{d\sigma^{(qN \rightarrow \gamma^* X)}}{ d \ln \alpha } \,,
\end{equation}
where
\begin{equation}
\label{eq-qN}
  \frac{d\sigma^{(qN \rightarrow \gamma^* X)}}{ d \ln \alpha} =
  \int d^2\rho
  |\Psi_{\gamma^*q}(\alpha,\vec\rho\,)|^2
  \sigma_{q \bar q}(\alpha\vec \rho\,) \,.
\end{equation}
The formula (\ref{eq-pp}) can be straightforwardly generalised to the DY process off nuclear targets by a replacement of the quark-nucleon DY cross section 
in Eq.~(\ref{eq-pp}) by a quark-nucleus one. 

The Green function formalism \cite{kst99} represents a universal method describing dipole interactions with nuclear targets in the whole kinematic region 
with no restrictions imposed on the CL. This formalism is also relevant for proper understanding of the GS effects emerging due to a contribution 
of higher Fock components containing gluons. In the latter, the dominating scales at large energies and, especially, at forward rapidities (or large Feynman $x_F$) 
are small such that the corresponding CL is comparable to the nuclear radius and thus has to be treated exactly.  

In the framework of Green function formalism, the quark-nucleus cross section can be written as a combination of two terms,
\begin{eqnarray}
\label{eq:dipole:qAGreen-pTint}
\frac{d\sigma^{(qA \to \gamma^* X)}}{d \ln \alpha}
&=& A \frac{d\sigma^{(qN \to \gamma^* X)}}{d \ln \alpha}
  -\frac{d\Delta\sigma^{(qA \to \gamma^* X)}}{d \ln \alpha} 
\nonumber\\
&=& A \frac{d\sigma^{(qN \to \gamma^* X)}}{d \ln \alpha}
  -\frac{1}{2}\textrm{Re} \int d^2b \int_{-\infty}^{\infty} dz_1 
\int_{z_1}^{\infty} dz_2 \int d^2\rho_1\,d^2\rho_2 
\nonumber\\
 &\times&  \Psi_{\gamma^*q}^\dagger (\alpha, \vec{\rho}_2) \rho_A(b,z_2) 
\sigma_{q \bar q}(\alpha \vec\rho_2) G(\vec{\rho}_2, z_2 | \vec{\rho}_1, z_1) 
\nonumber\\
 &\times&  \rho_A(b,z_1) \sigma_{q \bar q}(\alpha \vec\rho_1)  
\Psi_{\gamma^*q} (\alpha, \vec{\rho}_1)\,,
\end{eqnarray}
where the first term represents incoherent quark-nucleon interactions while the second term takes into account quark interactions with the nuclear target
and, thus, represents the shadowing correction of the lowest order. The quark Green function $G(\vec{\rho}_2, z_2 | \vec{\rho}_1, z_1)$ satisfies 
the two-dimensional Schr\"odinger equation
\begin{equation}
\label{eq:Dipole:pA:Green:SchoedingEq}
  \left[ i \frac{\partial}{\partial z_2} + \frac{\Delta(\vec{\rho}_2)-\eta^2}{2 E_q \alpha (1 - \alpha)} - 
  V(b, \vec{\rho}_2, z_2)  \right]
  G(\vec{\rho}_2, z_2 | \vec{\rho}_1, z_1) = 
  i \delta(z_2 - z_1) \delta^2(\vec{\rho}_2 - \vec{\rho}_1) \,,
\end{equation}
where $\eta^2 = (1 - \alpha)M_{l\bar{l}}^2 + \alpha^2 m_q^2$, and the second term on the l.h.s. represents the kinetic term which accounts 
for varying effective mass of the $q\gamma^*$ fluctuation and provides the corresponding phase shift. Note, the two-dimensional Laplacian in Eq.~(\ref{eq:Dipole:pA:Green:SchoedingEq}) acts on the transverse coordinate $\vec{\rho}_2$.

The imaginary part of the potential $V(b, \vec{\rho}_2, z_2)$ in the Schr\"odinger equation (\ref{eq:Dipole:pA:Green:SchoedingEq}) reads
\begin{equation}
\label{eq:Dipole:pA:Green:IMagPotential}
 {\rm Im}\, V(b, \vec{\rho}, z) = - \frac{1}{2} \rho_A(b, z) \sigma_{q\bar q}(\alpha\vec\rho\,) \,.
\end{equation}
This quantity is responsible for an attenuation of the $q\gamma^*$ fluctuation propagating in the medium since it is proportional to the nuclear density
function $\rho_A(b,z)$ which depends on the impact parameter $b$ and longitudinal coordinate $z$. This potential effectively accounts for all 
higher-order scattering terms as in the Glauber theory. One could notice an analogy with the optical theorem where the effect of absorption in the medium 
is also described by an imaginary potential. For convenience, the factor $\exp(-i q_L^{min}(z_2-z_1))$ describing the longitudinal motion is included 
into the Green function $G(\vec{\rho}_2, z_2 | \vec{\rho}_1, z_1)$ as was proposed in Ref.~\cite{Raufeisen:2000sy}.
\begin{figure}[t]
\large
\begin{center}
\scalebox{0.6}{\includegraphics{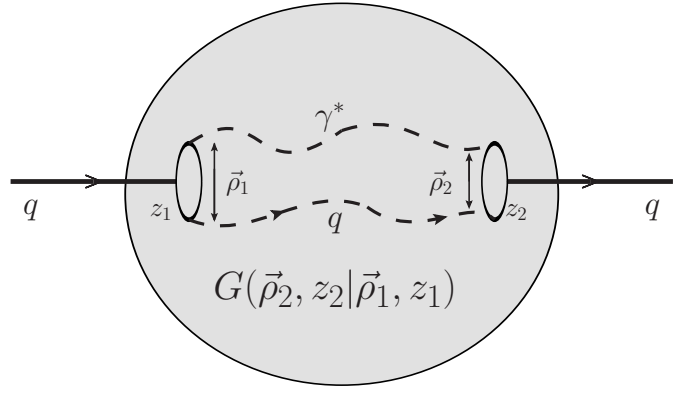}}
\caption{An illustration representing the nuclear correction (shadowing) term in Eq.~(\ref{eq:dipole:qAGreen-pTint}).
The propagation of the $q\gamma^*$ fluctuation through the nucleus is described by the Green function $G(\vec{\rho}_2, z_2 | \vec{\rho}_1, z_1)$
as a result of summation over different paths of the $q\gamma^*$ state in the medium.
}
\label{fig:dipole:green-qA}
\end{center}
\end{figure}
\normalsize

The second (shadowing) term in Eq.~(\ref{eq:dipole:qAGreen-pTint}) is illustrated in Fig.~\ref{fig:dipole:green-qA} and
can be interpreted as follows. At the point $z_1$ the projectile quark diffractively produces a $|q\gamma^*\rangle$ state 
($q N\rightarrow\gamma^* q N$) with an initial transverse separation $\vec{\rho}_1$. Then such a $|q\gamma^*\rangle$ fluctuation 
propagates through the nucleus along arbitrary curved paths, which are summed over, and arrives at the point $z_2$ with a 
transverse separation $\vec{\rho}_2$. The initial and final separations are controlled by the LC wave functions of the $|q\gamma^*\rangle$ 
Fock state in the projectile, namely, by $\Psi_{q\gamma^*}(\alpha,\vec{\rho})$. While propagating through the nucleus, the $|q\gamma^*\rangle$ 
state interacts with bound nucleons. The latter interaction is described by the dipole cross section, $\sigma_{q\bar q}(\alpha\vec{\rho})$, which depends 
on the local transverse separation $\vec{\rho}$ evolving over the propagation time scale. The Green function denoted as $G(\vec{\rho}_2, z_2 | \vec{\rho}_1, z_1)$
determines the evolution and properties of such a Fock state on its move from point $z_1$ to $z_2$.

If the high-energy limit $E_q \to \infty$ is concerned, the kinetic term in Eq.~(\ref{eq:Dipole:pA:Green:SchoedingEq}) can be neglected 
yielding the Green function in the following form,
\begin{equation}
\label{eq:Dipole:pA:Green:LCL}
  \left.G(\vec{\rho}_2, z_2 | \vec{\rho}_1, z_1)\right|_{E_q \to \infty} = 
  \delta^2(\vec{\rho}_1 - \vec{\rho}_2) 
  \exp \left[ i \int_{z_1}^{z_2}dz\,V(b,\vec{\rho}_2,z) \right]\,,
\end{equation}
thus demonstrating that the transverse separation of the $q\gamma^*$ fluctuation is frozen during its propagation through the nuclear medium.
After substitution of Eq.~(\ref{eq:Dipole:pA:Green:LCL}) into Eq.~(\ref{eq:dipole:qAGreen-pTint}) one obtains expressions for the DY nuclear 
production cross section in the LCL limit initially presented in Ref.~\cite{kst99}.

The analysis of the two-dimensional Schr\"odinger equation for arbitrary dipole cross section $\sigma_{q\bar q}(\alpha\rho)$ and a realistic nuclear 
density function $\rho_A(b, z)$ cannot be obtained analytically and should be performed using sophisticated numerical methods. The algorithm for 
numerical solution of this equation was proposed for the first time in Ref.~\cite{Nemchik:2003wx} and used for the Deep Inelastic Scattering (DIS) process. 
In what follows, we extend this numerical algorithm also for the DY process. For this purpose, it is convenient to rewrite Eq.~(\ref{eq:dipole:qAGreen-pTint}) 
eliminating the $\delta$-functions in Eq.~(\ref{eq:Dipole:pA:Green:SchoedingEq}). This can be done introducing the auxiliary functions,
\begin{eqnarray}
  g_1(\vec{\rho}_2, z_2 | z_1) 
&=& 
  \int d^2 \rho_1\, \textrm{K}_0(\eta \vec \rho_1) 
  \sigma_{q \bar q}(\alpha\vec \rho_1) G(\vec{\rho}_2,z_2|\vec{\rho}_1,z_1) \,, \\
  \frac{\vec{\rho}_2}{\rho_2} \, g_2(\vec{\rho}_2, z_2 | z_1) 
&=& 
  \int d^2\rho_1\, \textrm{K}_1(\eta \vec \rho_1) 
  \sigma_{q \bar q}(\alpha\vec \rho_1)\frac{\vec{\rho}_1}{\rho_1} 
  G(\vec{\rho}_2,z_2|\vec{\rho}_1,z_1) \,,
\end{eqnarray}
where $K_{0,1}$ are the modified Bessel functions of the second kind. The new Green functions $g_{1,2}$ 
satisfy the following evolution equations
\begin{eqnarray}
  i\frac{\partial}{\partial z_2}g_1(\vec{\rho}_2, z_2 | z_1) 
&=& 
  \left[ \frac{1}{2E_q\alpha(1-\alpha)} \left( \eta^2 - \frac{\partial^2}{\partial^2 \rho_2} - 
  \frac{1}{\rho_2}\frac{\partial}{\partial \rho_2} \right) + 
  V(z_2, \vec{\rho}_2, \alpha) \right]g_1(\vec{\rho}_2, z_2 | z_1) \,, 
\label{eq:Dipole:pA:DY:qA:exact-g1-eq} \\
  i\frac{\partial}{\partial z_2}g_2(\vec{\rho}_2, z_2 | z_1) 
&=& 
 \left[ \frac{1}{2E_q\alpha(1-\alpha)} \left( \eta^2 - \frac{\partial^2}{\partial^2 \rho_2} - 
 \frac{1}{\rho_2}\frac{\partial}{\partial \rho_2} + \frac{1}{\rho_2^2} \right) + 
 V(z_2, \vec{\rho}_2, \alpha) \right]g_2(\vec{\rho}_2, z_2 | z_1) 
\label{eq:Dipole:pA:DY:qA:exact-g2-eq}
\end{eqnarray}
with the boundary conditions defined as
\begin{eqnarray}
  g_1(\vec{\rho}_2, z_2 | z_1)|_{z_1=z_2} 
&=& 
  \textrm{K}_0(\eta \vec \rho_2) \sigma_{q \bar q}(\alpha\vec \rho_2) \,, 
\label{eq:Dipole:pA:DY:qA:exact-g1-BC} \\
  g_2(\vec{\rho}_2, z_2 | z_1)|_{z_1=z_2} 
&=& 
  \textrm{K}_1(\eta \vec \rho_2) \sigma_{q \bar q}(\alpha\vec \rho_2) \,.
\label{eq:Dipole:pA:DY:qA:exact-g2-BC}
\end{eqnarray}
Then the second term in Eq.~(\ref{eq:dipole:qAGreen-pTint}) can be conveniently expressed as follows,
\begin{eqnarray}
  \frac{d\Delta\sigma^{(qA \to \gamma^* X)}}{d\ln \alpha}
&=& 
  \alpha_{em} \textrm{Re} \int db\,b \int_{-\infty}^\infty dz_1 
  \int_{z_1}^\infty dz_2  \int d\rho_2\,\rho_2 
  \rho_A(b,z_1)\rho_A(b,z_2)\sigma_{q \bar q}(\alpha\vec \rho_2)  \nonumber\\
&\times & 
  \left[ (1+(1-\alpha)^2)\eta^2  \textrm{K}_1(\eta \vec \rho_2)g_2(\vec{\rho}_2, z_2 | z_1) \right.  \nonumber\\
&+& \left.
  (m_q^2 \alpha^4 + 2M_{l\bar{l}}^2(1-\alpha)^2) \textrm{K}_0(\eta 
  \vec \rho_2)g_1(\vec{\rho}_2, z_2 | z_1) \right] \,. 
\label{eq:Dipole:pA:DY:qA:Xsec-exact}
\end{eqnarray}
As was discussed in detail in Appendix~A of Ref.~\cite{Nemchik:2003wx}, the time-dependent two-dimensional Schr\"odinger equations 
given by Eqs.~(\ref{eq:Dipole:pA:DY:qA:exact-g1-eq}) and (\ref{eq:Dipole:pA:DY:qA:exact-g2-eq}) can be solved then by a modification 
of the method based on the Crank-Nicholson algorithm \cite{giordano1997computational,Goldberg:1967QM1D,crank1947}. The corresponding 
numerical results for the most important DY observables in the Green function framework are given and compared to those in the LCL/SCL regimes 
in the next Section.
\begin{figure}[t]
\large
\begin{center}
\scalebox{1.0}{\includegraphics{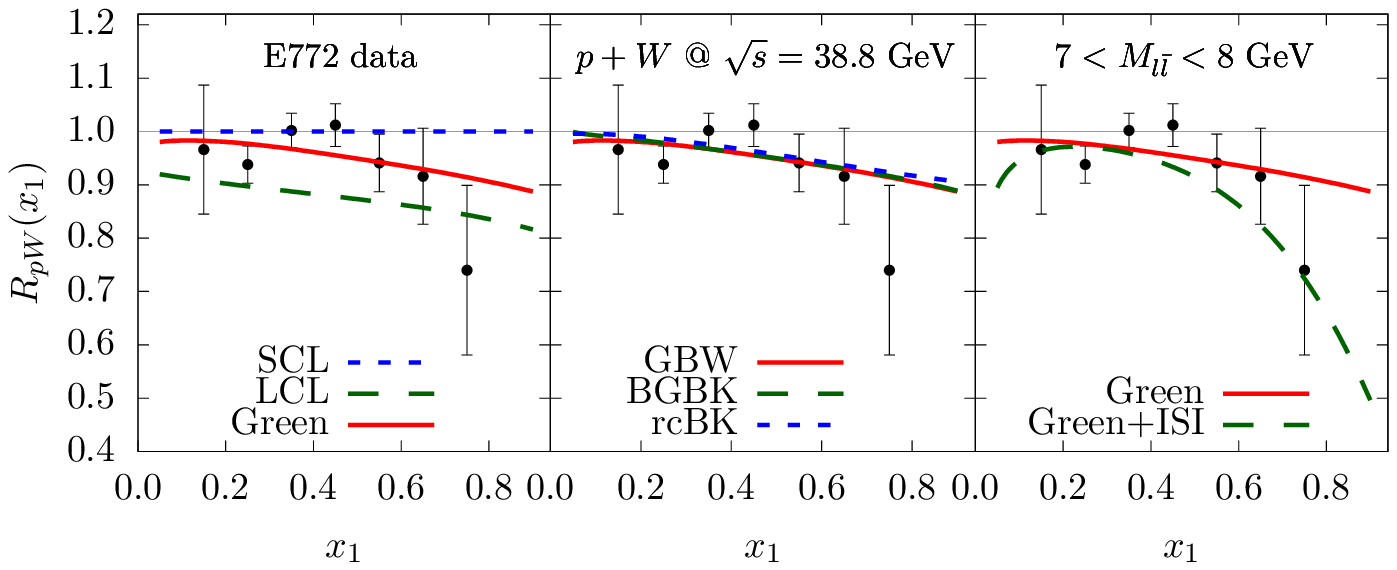}}
\scalebox{1.0}{\includegraphics{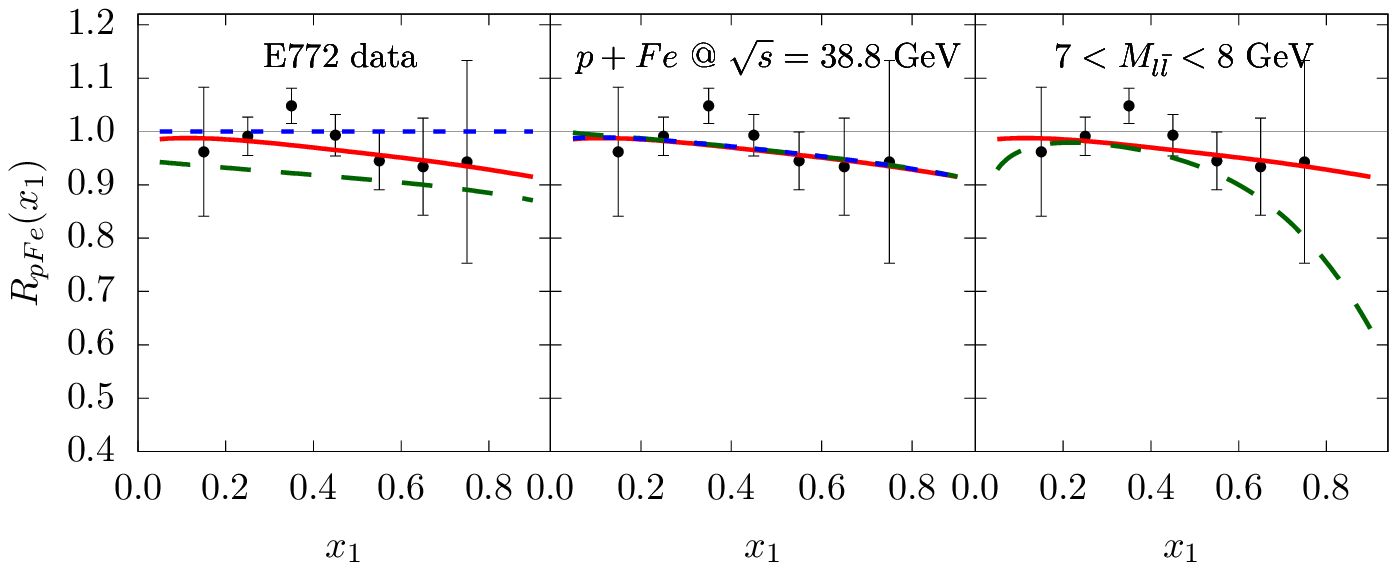}}
\caption{The nucleus-to-nucleon ratio $R_{pA}$ (nuclear modification factor) of the DY pair production cross sections in $pW$ (upper panels) 
and $pFe$ (lower panels) collisions as a function of the $x_1$ variable vs data from the E772 Collaboration \cite{E772}. The results are given
for the LCL and SCL approximations and are compared to those in the Green function approach in the left-most panels. Also, the predictions corresponding
to different models for the dipole cross section computed by using the Green function approach are presented in the middle panels. Finally, the results
of the Green function framework with and without an account for the initial-state interactions (ISI) are shown in the right-most panels.
}
\label{fig:fnal1}
\end{center}
\end{figure}
\normalsize

%
\section{Results}
\label{res}
%

In this Section, we present our predictions for the DY pair production cross sections in $pA$ collisions within the generalised color dipole approach based on the 
Green function formalism. The Green function $G(\vec{\rho}_2, z_2 | \vec{\rho}_1, z_1)$  is obtained as a result of the exact numerical solution of 
the two-dimensional Schr\"odinger equation (\ref{eq:Dipole:pA:Green:SchoedingEq}) that describes the propagation of the $q\gamma^*$ Fock state 
in the nuclear medium. In our calculations, we employ the MSTW parameterisation \cite{mstw2008} for PDFs in the incoming proton. In order to compare 
our predictions obtained with no restrictions to a magnitude of the CL with commonly used approximations, we present also the corresponding 
results derived in the SCL \cite{Johnson:2001xfa} and LCL \cite{BGKNP} regimes. Moreover, one should estimate theoretical uncertainties associated 
with a shape of the dipole cross section $\sigma_{q\bar q}(\alpha\vec \rho)$ which is one of the basic ingredients in our calculations (see e.g. 
Eq.~(\ref{eq:Dipole:pA:DY:qA:Xsec-exact})). For this purpose, we also present the results for DY nuclear cross section obtained with several distinct models 
for $\sigma_{q\bar q}(\alpha\vec \rho)$ proposed in Refs.~\cite{gbw,bgbk,bkrunning} whose parameters were extracted by fits to the DIS HERA data. 

The calculations are performed for three different parameterisations for the dipole cross section. One of them is based on the saturation model proposed 
in Ref.~\cite{bgbk} and is denoted as BGBK in what follows. Besides, we take into account also the dipole cross section model associated to a solution of 
the Balitsky-Kovchegov equation with the running coupling (rcBK) \cite{bk} which was obtained in Ref.~\cite{bkrunning}. Finally, as a naive reference model
we employ the phenomenological saturated parameterisation proposed by Golec-Biernat and W\"usthoff (GBW) in Ref.~\cite{gbw}. The later has already 
been used in our previous studies of the DY process. More details about these models can be found e.g. in Refs.~\cite{Basso_pp,BGKNP} as well as in original
articles mentioned above.
\begin{figure}[t]
\large
\begin{center}
\scalebox{1.0}{\includegraphics{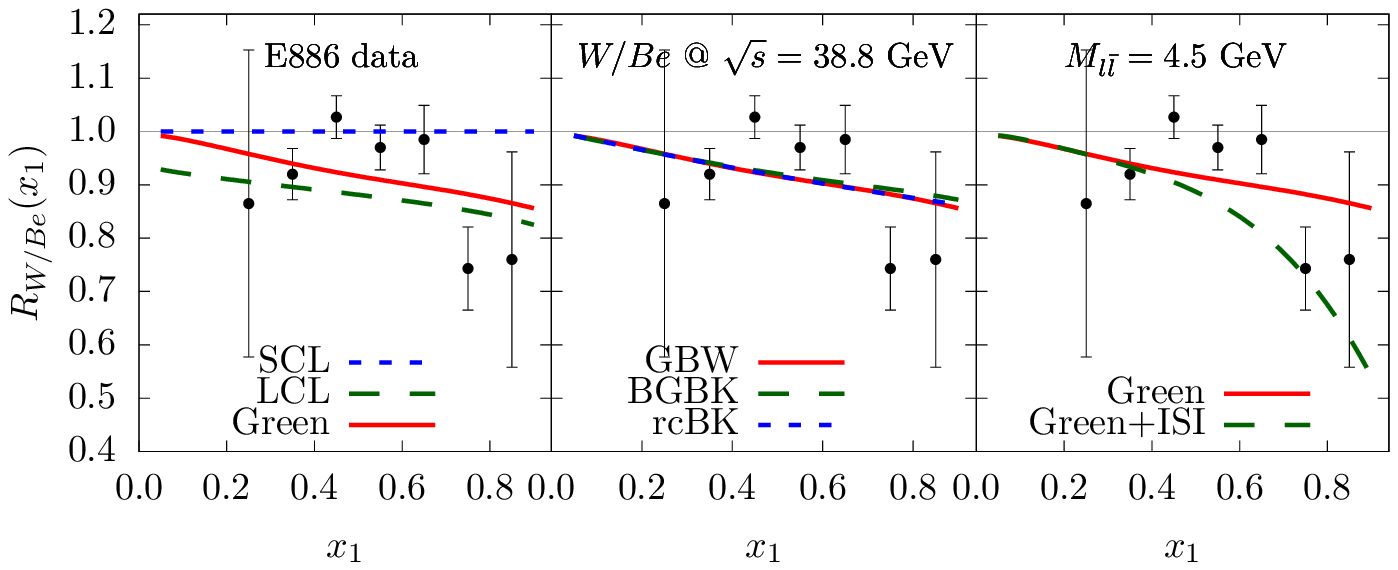}}
\scalebox{1.0}{\includegraphics{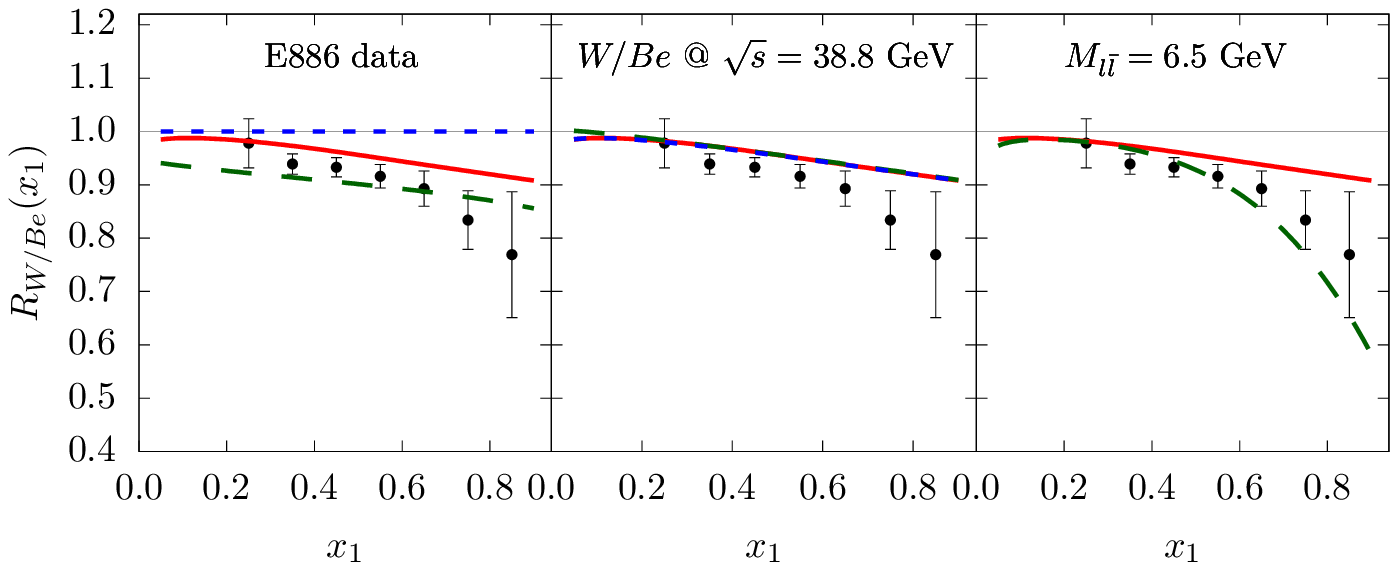}}
\caption{
The ratio of the DY pair production cross sections on tungsten (W) and beryllium (Be) targets $R_{W/Be}$ as a function of the $x_1$ variable
for the dilepton invariant mass 4.5 GeV  (upper panels) and 6.5 GeV (lower panels). The predictions are compared with the data from 
the E886 Collaboration \cite{E886}. Here, the meaning of the curves is same as in Fig.~\ref{fig:fnal1}.
}
\label{fig:fnal2}
\end{center}
\end{figure}
\normalsize

Besides the coherence effects controlled by the magnitude of the CL, in our study we also analyse the contribution of the initial-state energy loss effects 
due to initial state interactions (ISI) \cite{Kopeliovich:2005ym,Kopeliovich:2011zza} which may significantly affect the nuclear attenuation in DY pair production.
Similarly to the case of nuclear shadowing, we expect a stronger onset of ISI effects particularly at large Feynman $x_F\equiv x_L = 2p_L / \sqrt{s}$ (i.e. at 
forward rapidities) and/or at large $x_T = 2p_T / \sqrt{s}$. This means that one has to consider a mixture and interplay of the CL and ISI effects 
over a broad kinematic region. However, in contrast to the CL effects, the ISI effects should contribute at any energy as well as in those kinematic regions
where no significant coherence effects are expected. For this reason, when analysing various processes on nuclear targets, a clean study of the net ISI effect 
requires going to smaller energies where the CL effects are negligible. This then leads to significant restrictions on relevant kinematic regions which excludes 
those accessible at RHIC and LHC.

In comparison to other well-known processes on nuclear targets, the DY reaction is exceptionally useful for studies of the net ISI effect since it allows to eliminate 
the CL effects becoming a clean probe for the ISI effects at any energy simply by e.g. turning to large values of $x_1$ (or Feynman $x_F$). Indeed, according 
to Eq.~(\ref{eq-cl}) the CL vanishes in the limit $x_1,\, x_F,\, \alpha \rightarrow 1$ approaching the SCL regime at $x_F>0.9$ as is shown in Fig.~\ref{fig:lc_xF}. 
Another way to eliminate the coherence effects is to go to larger values of the dilepton invariant mass $M_{l\bar l}$ as is demonstrated in Figs.~\ref{fig:lc_M} 
and \ref{fig:lc_xF}.

The ISI effects were studied in Refs.~\cite{Kopeliovich:2005ym,Kopeliovich:2011zza} within the Glauber approximation where each interaction in the nucleus leads 
to a suppression factor $S(\xi) \approx 1 - \xi$, where $\xi = \sqrt{x_L^2 + x_T^2}$. The summation over multiple ISIs at the impact parameter $b$ leads 
to the following nuclear ISI-modified PDF,
\begin{equation}
\label{eq:qcd:isi}
  f_{q}(x,Q^2) \, \Rightarrow \, f_{q}^A(x,Q^2,b) = 
  C_v f_{q}(x,Q^2) \frac{e^{-\xi \sigma_{\rm eff}T_A(b)} - 
  e^{-\sigma_{\rm eff}T_A(b)}}{(1-\xi)(1-e^{-\sigma_{\rm eff}T_A(b)})}\,,
\end{equation}
where $\sigma_{\rm eff} = 20$~mb \cite{Kopeliovich:2005ym} is the hadronic cross section which effectively determines the rate of multiple interactions, and 
the normalisation factor $C_v$ is fixed by the Gottfried sum rule.

Additionally, above $\sqrt{s} = 100$ GeV we take into account an extra contribution to the nuclear shadowing coming from the higher Fock fluctuations containing 
gluons, the so-called GS effect. The latter is effectively incorporated into the calculations by a replacement, $\sigma_{q \bar q}(\vec\rho,x) \rightarrow 
\sigma_{q \bar q}(\vec\rho, x)\,R_G(x,Q^2,\vec b)$, where the corresponding suppression factor $R_G$ was derived in Ref.~\cite{kopeliovich-gs} in the framework of 
Green function technique. 
\begin{figure}[t]
\large
\begin{center}
\scalebox{1.0}{\includegraphics{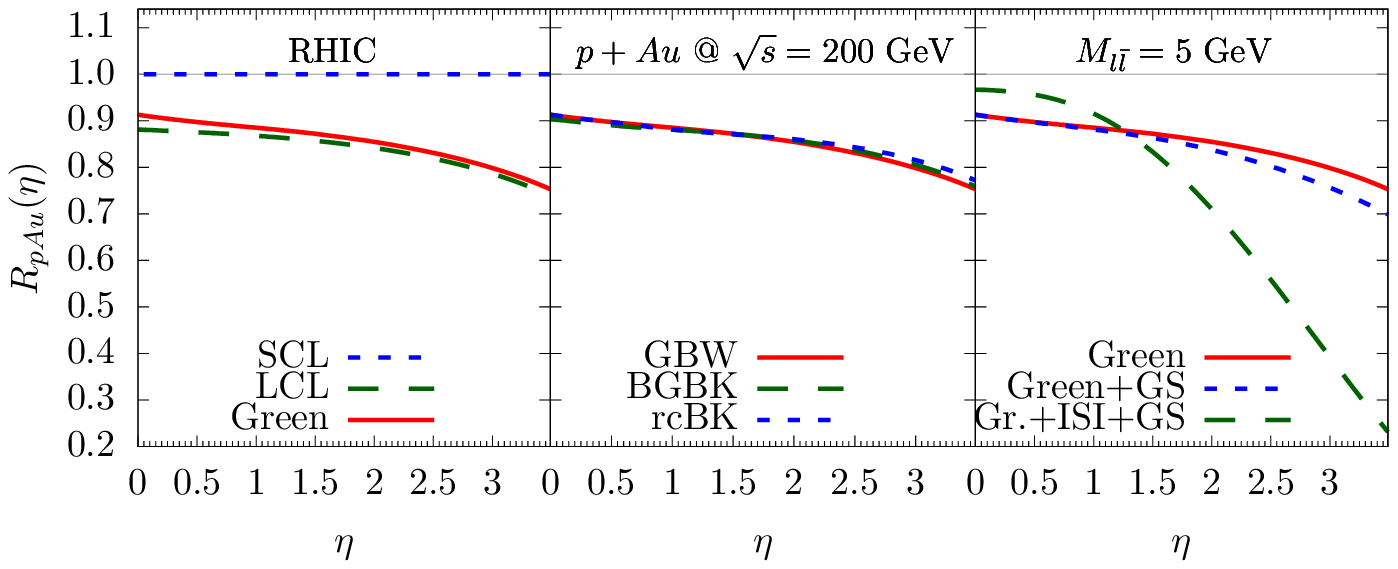}}
\scalebox{1.0}{\includegraphics{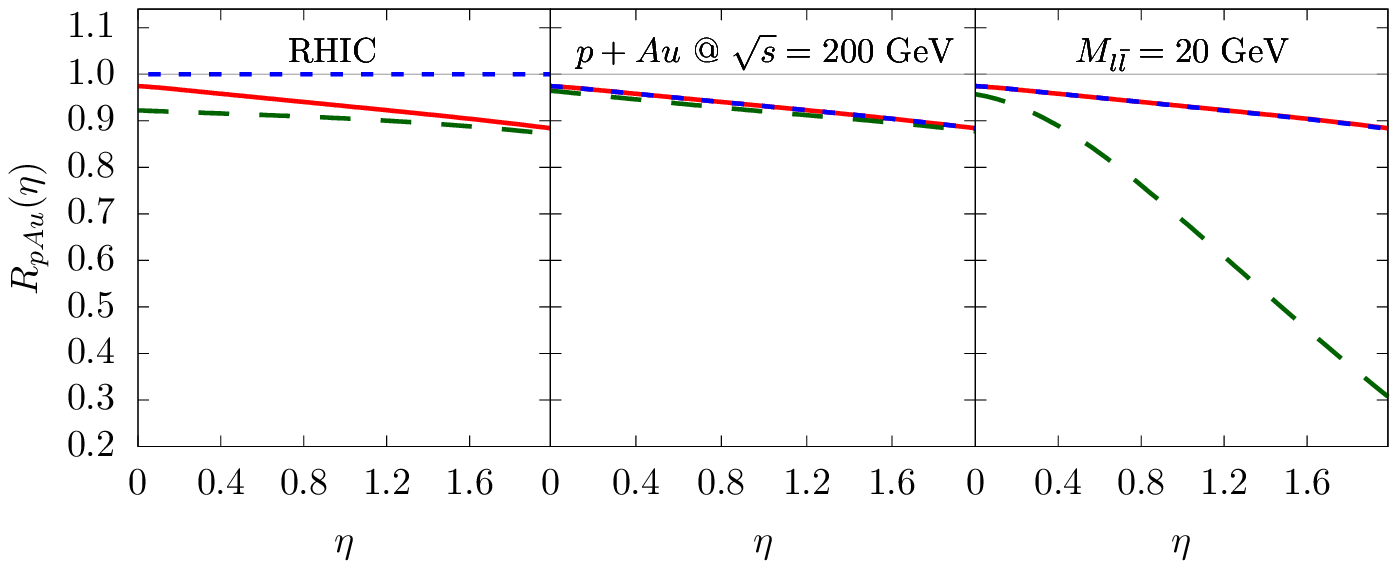}}
\caption{
The pseudorapidity dependence of the $R_{pAu}$ ratio at RHIC c.m. energy $\sqrt{s} = 200$ GeV and two different values of the dilepton invariant mass
$M_{l\bar l} = 5$ GeV (upper panels) and 20 GeV (lower panels). Here, the meaning of the curves is same as in Fig.~\ref{fig:fnal1} except that two additional
curves representing the GS effect are shown in the right-most panels, for comparison.}
\label{fig:mass_rhic1}
\end{center}
\end{figure}
\normalsize
\begin{figure}[t]
\large
\begin{center}
\scalebox{1.0}{\includegraphics{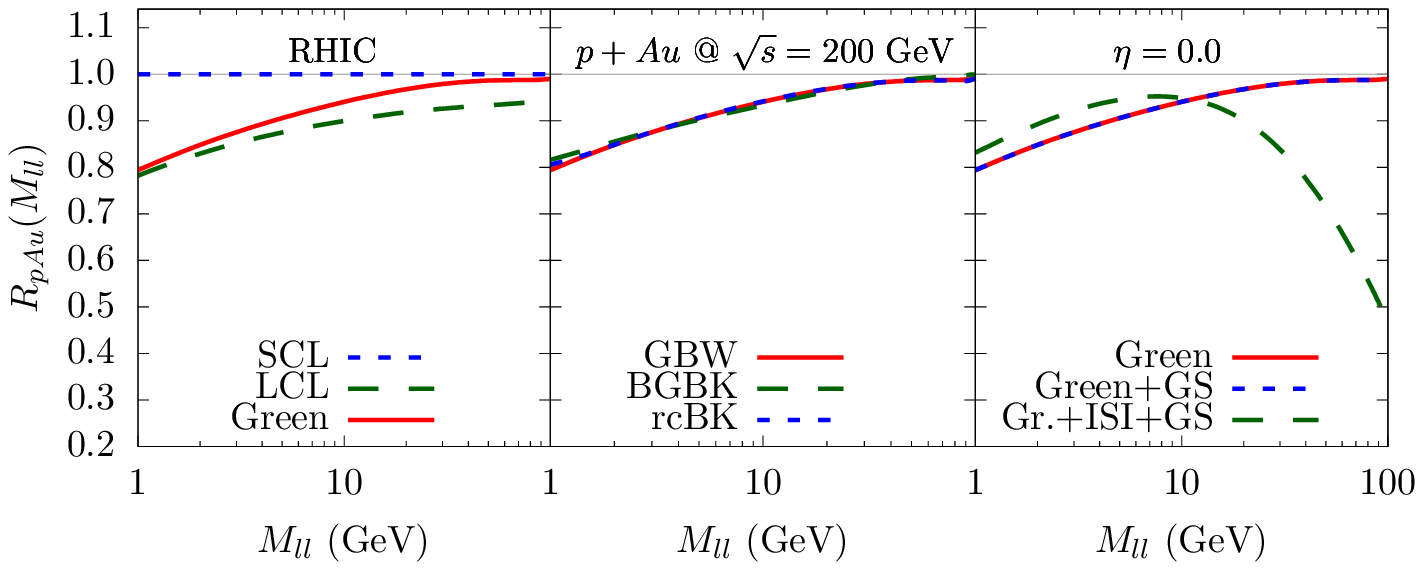}}
\scalebox{1.0}{\includegraphics{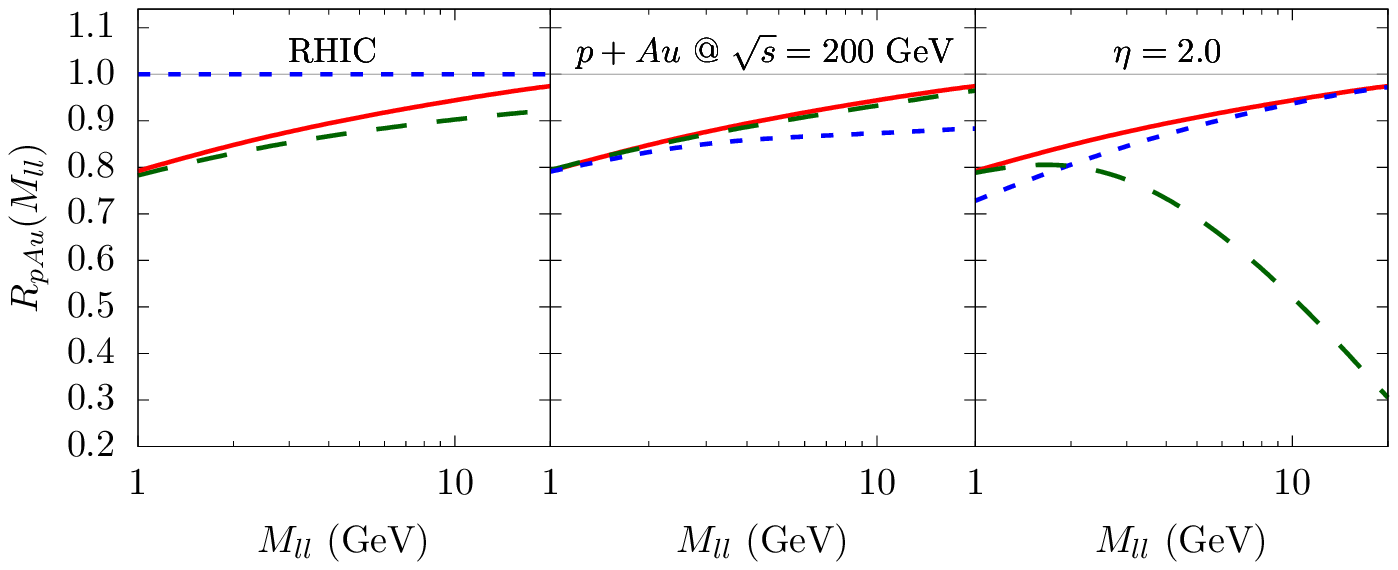}}
\caption{
The dilepton invariant mass dependence of the ratio $R_{pAu}$ at RHIC c.m. energy $\sqrt{s} = 200$ GeV and two different values of the pseudorapidity 
$\eta = 0$ (upper panels) and 2 (lower panels).  Here, the meaning of the curves is same as in Fig.~\ref{fig:mass_rhic1}.}
\label{fig:mass_rhic2}
\end{center}
\end{figure}
\normalsize

We start our analysis of the DY process on nuclear targets testing the predictions against available data from the fixed-target E772 and E886 experiments at FNAL
\cite{E772,E886}. Such a comparison can be performed for different nuclei and several values of the dilepton invariant mass. In Fig.~\ref{fig:fnal1} we present results
for the nuclear modification factor $R_{pA}$ as a function of $x_1$ for tungsten (upper panels) and iron (lower panels) targets vs the E772 data \cite{E772}.
In the left-most panels, we compare the rigorous Green function formalism with the results obtained in the SCL and LCL limits and find that in the considered
kinematic region the onset of CL effects rises gradually with $x_1$ (or, equivalently, with the Feynman $x_F$ and rapidity). This is seen from an increase of 
the difference between dotted and dashed curves corresponding to the SCL (with no CL effects included) and LCL (with maximal CL effects) limits, respectively.
Since the nuclear radii satisfy $R_{Fe} < R_{W}$, we obtain a smaller shadowing effect for the iron target as is shown in lower panels of Fig.~\ref{fig:fnal1} as expected. 

Here, it is worth emphasizing that calculations performed in the LCL limit and often presented in the literature lead to a rather large overestimation of the nuclear 
shadowing in comparison with the Green function formalism (see e.g. differences between dashed and solid lines in left panels of Fig.~\ref{fig:fnal1}. Thus, one 
concludes that the LCL limit should not be used in analysis of the nuclear shadowing in DY pair production in kinematic regions corresponding to the fixed-target
FNAL experiments. 

The results for $R_{pA}(x_1)$ presented in middle panels of Fig.~\ref{fig:fnal1} clearly demonstrate that predictions obtained in the framework of 
Green function formalism are practically independent of the shape of the dipole cross section (at least, for parameterisations used in our analysis). Finally, the results on 
right-most panels of Fig.~\ref{fig:fnal1} show a strong onset of ISI effects that significantly rises with $x_1$ and leads to a better agreement of our predictions 
with the E772 data.

Analogous conclusions can be drawn from Fig.~\ref{fig:fnal2} where the ratio $R_{W/Be}$ is given as a function of $x_1$-variable for two different values 
of the dilepton invariant mass $M_{l\bar{l}} = 4.5$ GeV (upper panels) and 6.5 GeV (lower panels). The results have been compared to the fixed-target 
FNAL E886 data \cite{E886}. One immediately notices that the onset of ISI effects turns out to be much stronger than the nuclear shadowing
at large $x_1$. This means that the kinematical region corresponding to the fixed-target FNAL experiment is very convenient for probing the ISI effects.

In Figs.~\ref{fig:mass_rhic1} and \ref{fig:mass_rhic2} we show our predictions for the nuclear effects in the DY pair production off the gold target 
in the kinematic region accessible by the RHIC collider. Here, we observe a stronger onset of coherence effects than that found for the fixed-target FNAL 
c.m. collision energy $\sqrt{s} = 38.8$ GeV. This is demonstrated in Fig.~\ref{fig:mass_rhic1} showing the pseudorapidity dependence of 
the nuclear ratio $R_{pAu}$ for two different values of the dilepton invariant mass $M_{l\bar l} = 5$ GeV and 20 GeV. One notices in left-most panels
of Fig.~\ref{fig:mass_rhic1} that calculations performed in the LCL limit and in the framework of Green function formalism yield very similar results
for the nuclear shadowing, in particular, at large $\eta$ and for a smaller value of $M_{l\bar l} = 5$ GeV since the CL exceeds the nuclear radius in this case.
However, the magnitude of the CL decreases with increasing dilepton invariant mass such that the LCL approximation breaks down and can not be used
for calculation of the nuclear shadowing. This is shown in the lower left panel of Fig.~\ref{fig:mass_rhic1} where one notices a sizeable deviation of the Green 
function formalism predictions from those in the LCL approximation.

Besides, theoretical uncertainties in the nuclear shadowing were tested by using various models for the dipole cross section $\sigma_{q\bar{q}}$. 
As is shown in the middle panels of Fig.~\ref{fig:mass_rhic1}, there is only a minor difference between the corresponding results. This observation 
stems from the fact that various dipole parameterisations used in our analysis are probed in the range of transverse separations where they have 
a rather similar $\rho$-behaviour.

At smaller $M_{l\bar l} = 5$ GeV and at $\eta\geq 2$ shown in the right-most panels of Fig.~\ref{fig:mass_rhic1}, a minor correction to the nuclear 
shadowing comes from the GS effect leading to an additional suppression. Indeed, this was expected since the CL corresponding to the $|q\gamma^*G\rangle$
fluctuation becomes comparable to the nuclear radius. At larger $M_{l\bar l} = 20$ GeV the GS correction turns out to be negligible. As was already observed 
for the fixed-target FNAL energy, in addition to the shadowing effects, the ISI effects play a dominant role at $\sqrt{s}=200$ GeV as well (especially, at large 
rapidities and dilepton invariant masses) and cause an extra significant suppression as is shown in the right-most panels of Figs.~\ref{fig:mass_rhic1} 
and \ref{fig:mass_rhic2}. The left-most panels of Fig.~\ref{fig:mass_rhic2} clearly demonstrate a gradual elimination of CL effects with an increase of $M_{l\bar l}$.
The upper right panel of Fig.~\ref{fig:mass_rhic2} shows that the CL effects are almost entirely absent at $M_{l\bar l}\geq 50$ GeV and $\eta = 0$
such that the predictions for the ratio $R_{pAu}(M_{l\bar l})$ represent the onset of ISI effects. Here, one should emphasize that the LCL approximation can 
not be valid anymore at large $M_{l\bar l}\geq 5\div 10$ GeV, depending also on rapidity. Finally, in lower middle panel of Fig.~\ref{fig:mass_rhic2} 
we demonstrate that the rcBK model for the dipole cross section leads to the results which deviate from those obtained by using the GBW and BGBK models, 
especially at larger pseudorapidities and dilepton invariant masses. 
\begin{figure}[t]
\large
\begin{center}
\scalebox{1.0}{\includegraphics{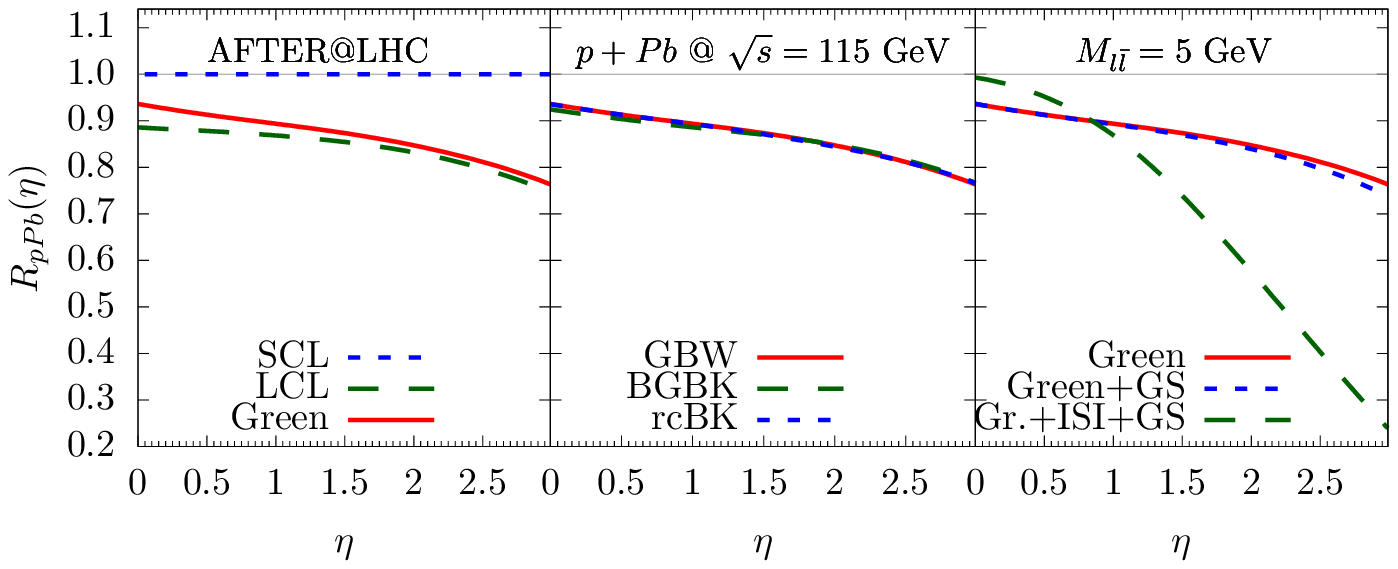}}
\scalebox{1.0}{\includegraphics{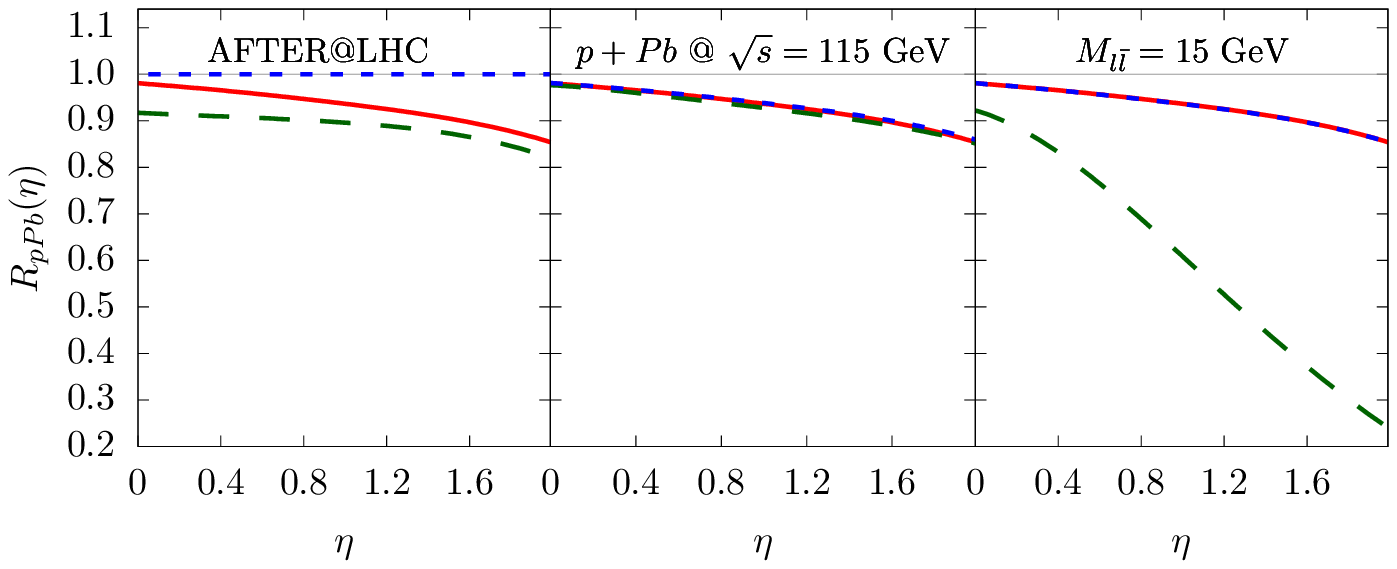}}
\caption{
The pseudorapidity dependence of the lead-to-proton ratio $R_{pPb}$ at the AFTER@LHC c.m. collision energy $\sqrt{s} = 115$ GeV and
for two different values of the dilepton invariant masses $M_{l\bar l} = 5$ and 15 GeV.
}
\label{fig:after1}
\end{center}
\end{figure}
\normalsize
\begin{figure}[t]
\large
\begin{center}
\scalebox{1.0}{\includegraphics{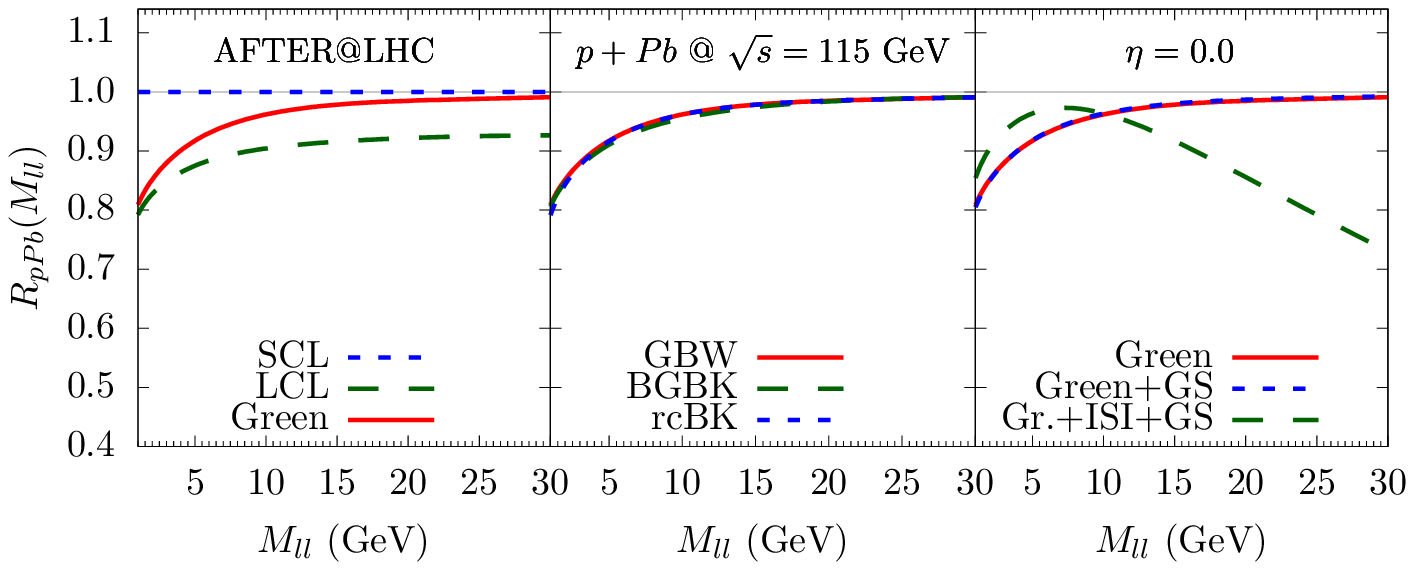}}
\scalebox{1.0}{\includegraphics{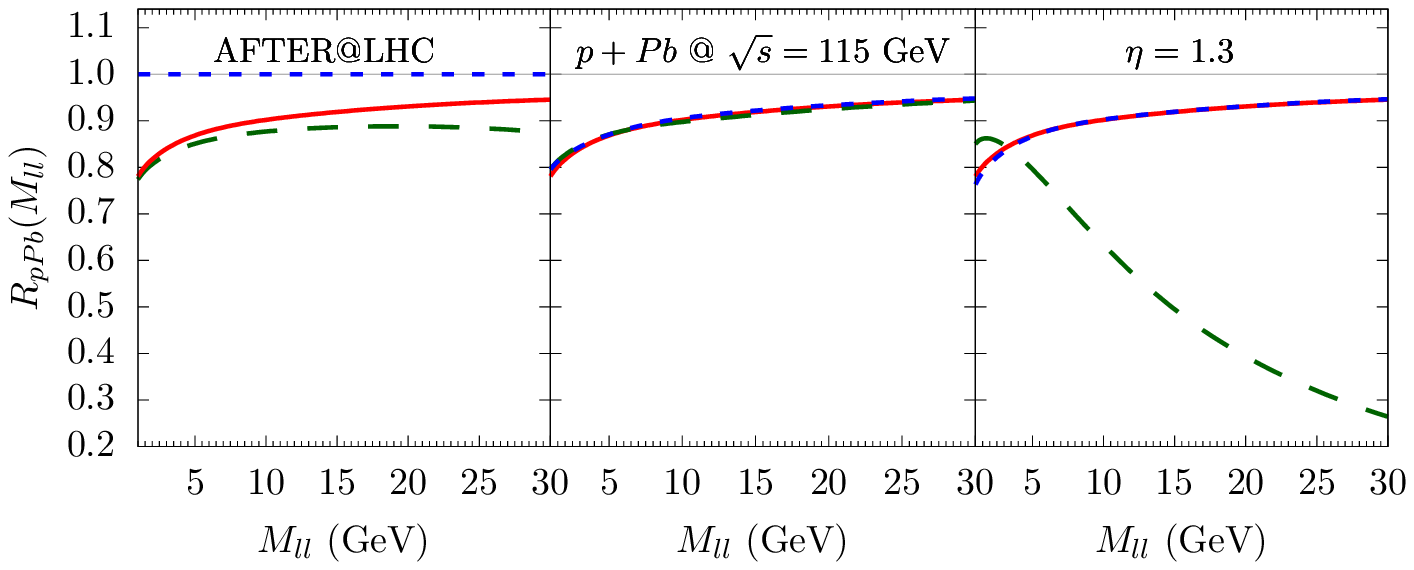}}
\caption{
The dilepton invariant mass dependence of the lead-to-proton ratio $R_{pPb}$ at AFTER@LHC c.m. collision energy $\sqrt{s} = 115$ GeV and 
for two different values of the pseudorapidity $\eta = 0$ and 1.3.
}
\label{fig:after2}
\end{center}
\end{figure}
\normalsize

The proposal for a fixed-target experiment at the LHC \cite{after} known as AFTER@LHC as well as the recent studies of 
proton-gas fixed-target collisions \cite{lhcb_fixedtarget} by the LHCb Collaboration strongly motivate us to make an extension 
of the above predictions to the nuclear targets and kinematic regions covered by these experiments.

In Figs.~\ref{fig:after1} and \ref{fig:after2} we present predictions for the nuclear attenuation in DY pair production 
that can be measured by the AFTER@LHC experiment in $pPb$ collisions at c.m. collision energy $\sqrt{s} = 115$ GeV. 
These clearly demonstrate that the GS corrections to the nuclear shadowing are rather small and can be neglected. 
On the other hand, one observes a strong onset of the ISI effects at forward rapidities and/or at large $M_{l\bar{l}}$ where 
only a weak effect of the quantum coherence is expected.

\begin{figure}[t]
\large
\begin{center}
\scalebox{1.0}{\includegraphics{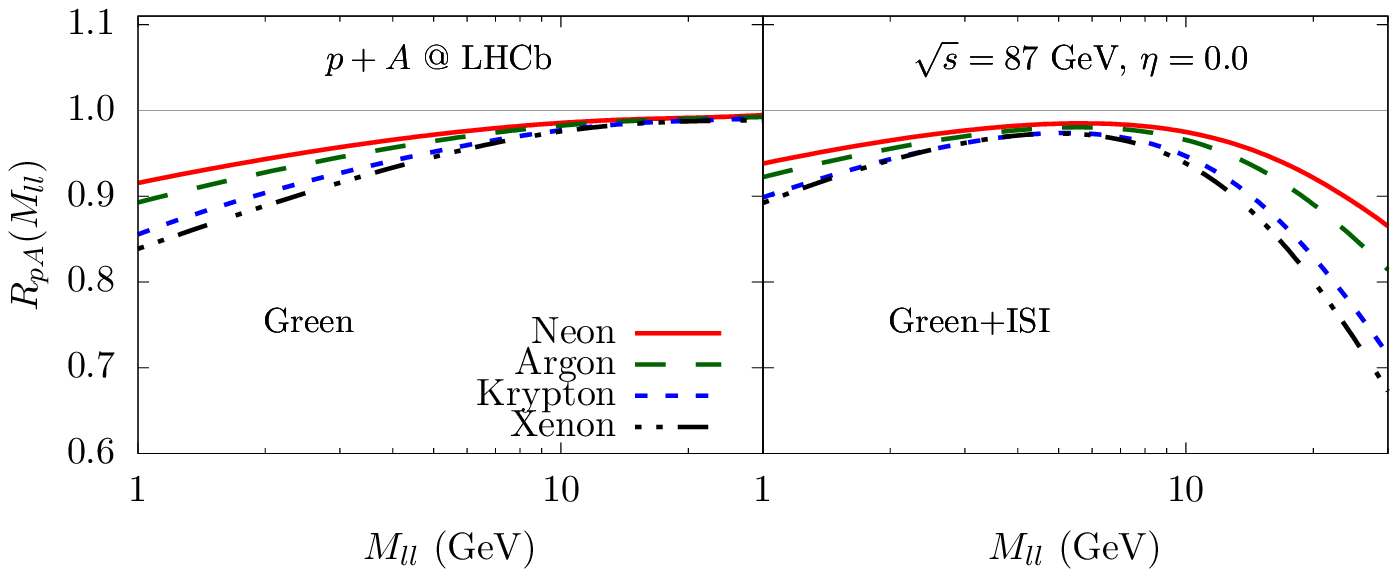}}
\scalebox{1.0}{\includegraphics{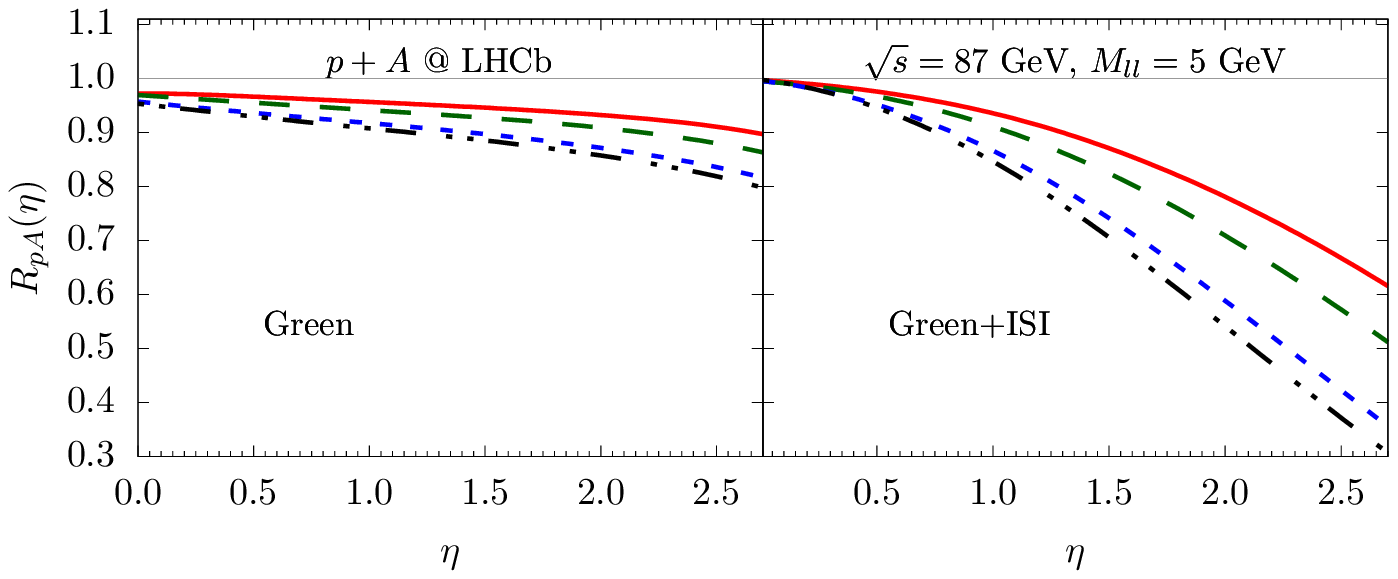}}
\caption{
The dilepton invariant mass (upper panels) and pseudorapidity (lower panels) dependences of the nucleus-to-proton 
ratio $R_{pA}$ at LHCb fixed-target c.m. collision energy $\sqrt{s} = 87$ GeV for several different nuclear targets.
}
\label{fig:mass_lhcb}
\end{center}
\end{figure}

Finally, in Fig.~\ref{fig:mass_lhcb} we show our results for the dilepton invariant mass and pseudorapidity 
dependence of the ratio $R_{pA}$ at LHCb fixed-target c.m. collision energy $\sqrt{s} = 87$ GeV considering several different 
targets. We observe an increase of the nuclear shadowing and/or ISI effects with increasing rapidity as expected. 
As was already indicated above, a nuclear DY measurement at large dilepton invariant masses would allow to suppress 
the CL effects and thus represents a clean promising way of phenomenological investigation of the net ISI effect.
\normalsize

%
\section{Summary}
\label{conc}
%

As was demonstrated by various studies during past few decades, the DY reaction in $pA$ collisions is an effective tool for clean studies 
of the initial-state medium effects occurring before a hard collision in the nuclear environment since no final-state interactions 
are concerned. In this work, we have shown that the relative contribution of nuclear effects to the nuclear suppression $R_{pA}$
is controlled by the CL which is correlated with the nuclear shadowing and depends on energy, rapidity and dilepton invariant mass. 

In the range of small collision energies when the CL is small, namely, $\l_c\leq 1\div 2$ fm, one should not expect any shadowing effects 
(SCL regime) due to a short lifetime of the $|q\gamma^*\rangle$ fluctuation. In contract, at large collision energies and/or forward rapidities 
when the CL considerably exceeds the nuclear radius, i.e. $l_c\gg R_A$, within an extended kinematic region we expect a maximal 
shadowing (LCL regime). In both cases, the theoretical description of the DY process is significantly simplified and is widely known 
in the literature.

However, at medium c.m. collision energies, $\sqrt{s} \lesssim 200$ GeV, and/or at large dilepton invariant masses, the CL can be smaller 
or comparable with the nuclear radius, $l_c \lesssim  R_A$, and thereby corresponds to a transition region between the SCL and LCL limits.
In this case, calculations performed within the LCL approximation lead to a sizeable overestimation of the nuclear shadowing and, consequently, 
should not be used in analysis of the medium effects. For this reason, we employ the rigorous path-integral technique known as 
the Green function formalism implying no restrictions to the CL. This formalism is known to provide an exact treatment of shadowing effects 
in various kinematic regions. Such a generic study of nuclear shadowing in the DY process for arbitrary CLs is the main purpose of our work.

We verified that the Green function formalism successfully reproduces the SCL and LCL predictions for the expected nuclear attenuation 
in DY pair production off nuclei in the corresponding kinematic regimes. Besides, we took into account additional shadowing corrections 
which come from the higher Fock fluctuations containing gluons (GS effect). The latter become relevant at large energies and at forward 
rapidities when the corresponding CL for the higher $|q\gamma^*G\rangle$ Fock state is comparable or exceeds the nuclear radius. 

Besides the shadowing effects, we included also the ISI effects affecting considerably the nucleus-to-proton ratio $R_{pA}$ at forward 
rapidities (or large Feynman $x_F$). Including the shadowing and ISI effects and using the Green function technique, we obtained a rather 
good description of available data from the fixed-target FNAL E772 and E886 experiments for the nuclear modification factor as a function 
of the $x_1$-variable and dilepton invariant mass. We have also made the corresponding predictions at RHIC energy $\sqrt{s}=200$ GeV.
Finally, we considered the kinematic regions expected to be probed by the planned AFTER@LHC experiment as well as by the LHCb Collaboration 
in recent studies of fixed-target proton-gas collisions. In all these cases, we have analysed the nuclear shadowing and ISI effects as the main
sources of nuclear suppression of DY pairs that can possibly be verified by future RHIC and LHC measurements. Finally, we would like to point out 
that the DY process off nuclei is especially convenient for a clean phenomenological study of the ISI effects since for this process the nuclear coherence 
can be consistently eliminated by turning to large values of the dilepton invariant mass where the net ISI effect is pronounced. This is in contrast to other 
high-energy reactions on nuclear targets where one can only probe the CL-ISI mixing.

%
\section*{Acknowledgements}
%

V.P.G. has been supported by CNPq, CAPES and FAPERGS, Brazil.
R.P. is supported by the Swedish Research Council, contract number 621-2013-428.
J.N. and M.K. are partially supported by the grant 13-20841S of the Czech Science Foundation
(GA\v CR) and by the Grant M\v SMT LG15001. J.N. is supported by the Slovak Research and
Development Agency APVV-0050-11 and by the Slovak Funding Agency, Grant 2/0020/14.

%
\bibliographystyle{unsrt}

\end{document}